# Value Entropy Model: Metric Method of Service Ecosystem Evolution

Xiao Xue, Zhaojie Chen, Shufang Wang, Zhiyong Feng, Yucong Duan, Zhangbing Zhou

*Abstract*—With the development of cloud computing, service computing, IoT(Internet of Things) and mobile Internet, the diversity and sociality of services are increasingly apparent. To meet the customized user demands, a complex collaboration network should be formed with various IT services through cross-border integration. Under this context, various service ecosystems begins to emerge, such as software service ecosystem, cloud manufacturing service system and E-commerce service ecosystem, etc. However, service ecosystem is a complex social-technology system, which is characterized by complexity, integration, dynamics, territoriality, etc. Hence, how to measure and evaluate the evolving characteristics of service ecosystem is of great significance to promote its sound development. Based on this, this paper proposes a value entropy model to analyze and measure the performance of service ecosystem from the perspective of value network, which is conducive to find the optimized operation strategy of service ecosystem. In addition, a computational experiment system is established to verify the effectiveness of value entropy model, which stimulates the competitive evolution process of two service ecosystems with different strategies. The result shows that our model can provide new means and ideas for the analysis of service ecosystem evolution, and can also provide decision support for the intervention strategy selection.

*Index Terms*—Service ecosystem, Value entropy model, Evolution measurement, Computational experiment.

## I. INTRODUCTION

With the development of information technologies such as service science [1], cloud computing [2], Internet-ware [3] and mobile Internet, more and more enterprises and organizations encapsulate their business capabilities (e.g., resource, platform, software, business and data) into services (e.g., Web service, RESTFul service, OpenAPI and Mobile APP), and support dynamic composition and collaboration of services through service-oriented technologies such as Workflow, Composition/Mashup and Personalized Service. . These cross-organization services can meet the complex and changeable demands through dynamic integration and collaboration. In the long-term competition and cooperation, a complicated interactive relationship and dynamical collaboration among service nodes can be formed through their self-organization mechanism. Under the context of the rapid development of service-based economy [4] and software service technologies[5], service ecosystem is generated, which is featured by rapid growth, dynamic change, mutual correlation and self-adaption[6-9].

Service ecosystem is a complex socio-technical system. As shown in Fig.1, there are three main roles in service ecosystem, namely service providers, service consumers, and service operators. Service providers refer to those who are in possession of resources, and provide services to service consumers within a specific time. Service consumers refer to those who consume the resources. The supply and demand matching between service providers and service consumers is realized by service operators, thus creating and producing values. Its operation process is driven by various factors such as independent service node evolution, changing user demand and dynamic service collaboration relationship, with a distinct feedback mechanism [10,11].

The current global market is rapidly changing and user needs are increasingly individualized. Service providers often adapt to these changes through reforming and transforming their organizational mode to remain competitive.

Up to this day, service ecosystem has become an important factor in the fierce global market competition. Thus, how to effectively measure and promote the evolution of service ecosystem has become critical. However, due to the complexity of service ecosystem, its analysis is facing the following challenges:

- **Individual complexity**: In a service ecosystem, service providers are social, which increases the diversity, uncertainty and dynamics of service provision. At the same time, individuals with strong independent decision-making ability and adaptability are capable of continuous self-regulation and dynamic evolution based on environmental changes. As a result, the entire service ecosystem is no longer static and fixed, but always dynamically changing.
- **Organizational complexity**: In a service ecosystem, all service providers need to benefit from collaboration on the premise of not losing their flexibility, that is, there are conflicts between organization collaboration and individual autonomy. Hence, the stability of organizational structure has a relative property. The failure in local service nodes may give rise to cascade effect of the service network, and finally causes instability of the whole system. This further increases the complexity of system analysis.
- **Social complexity**: In a service ecosystem, every service

Thanks for the support provided by National Key Research and Development Program of China (No. 2017YFB1401200), National Natural Science Foundation of China (No.61972276, No. 61832014, No. 41701133). (Corresponding author: Xiao Xue and Shufang Wang)

Xiao Xue is with College of Intelligence and Computing, Tianjin University, also adjunct professor in the School of Computer Science and Technology, Henan Polytechnic University, P.R.China.  (E-mail: jzxuexiao@tju.edu.cn).

Shufang Wang is with School of Geographic and Environmental Sciences, Tianjin Normal University, P.R.China. (E-mail: sfwang@tjnu.edu.cn).

node has its specific function and location, and different composition forms can be utilized between nodes to fulfill complex demands. With the enhanced integration, composition and interoperability of services, the topological network between services is increasingly complex. Affected by this, internal system change or external intervention factors may cause unpredictable emergencies, making it hard to analyze and predict the system evolution path.

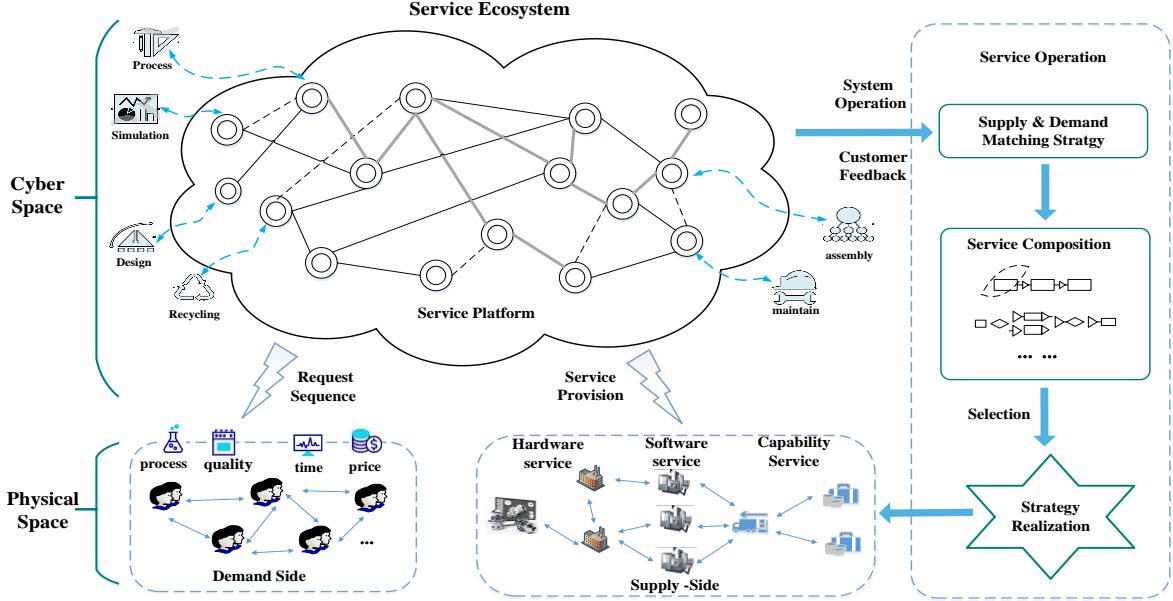

Fig.1 Operation diagram of service ecosystem

Due to the complexity of service ecosystem, the traditional Quality of Service [12] and System Performance Evaluation [13] are insufficient in the measurement of evolution status of service ecosystem. We need a method that can systematically analyze and evaluate the service ecosystem. In the field of thermodynamic, entropy is used to describe the chaos degree of a system. In the field of information, Shannon uses information entropy to describe the uncertainty of information sources. In the field of ecology, Shannon-Weiner index based on entropy theory is used to measure population diversity[14]. In service ecosystem, the orderliness of cooperation between service nodes directly determines the efficiency of value creation. Inspired by these ideas, we have proposed the value entropy model to measure the orderliness of service ecosystem and then evaluate the efficiency of the whole ecosystem in value creation and production.

The rest parts of this paper are organizaed as follows. Section II introduces relevant work of service ecosystem; Section III proposes the value entropy model, including entropy measurement, value analysis and operation strategy; Section IV designs the computational experiment system of service ecosystem from the perspective of supply and demand matching; Section V verifies the applicability of the value entropy model with an case study; Section VI discusses the effectiveness of value entropy model in practical cases; Section VII concludes the paper.

## II. BACKGROUND AND MOTIVATION

The concept of service ecosystem is originated from the ecosystem theory in ecology. Moore firstly applied the ecosystem thought in the business field and thereby proposed the concept of business ecosystem[15]. Subsequently, Vargo and Lusch proposed service-dominant logic to replace traditional commodity-dominant logic, defined the service ecosystem as a socio-technical system featured by complexity, self-evolution and autonomy [16]. The current researches on service ecosystem are mainly carried out from two aspects: analysis methods and specific domains. Its research status is as follows.

### A. Analysis method of service ecosystem

The analysis of the service ecosystem has always been the focus of academic circles, and its research is mainly divided into three parts:

**(1) Measurement and Evaluation**

To analyze service ecosystem, some researches analyze the scale, availability and complexity of services and other performance indexes (e.g., service round-trip time, throughput and utilization) by using the statistical analysis method. Masri and Mahmoud compared and analyzed the scale, availability and complexity of the obtained Internet services [17]. Zheng et al. collected 21,197 public services from the Internet and analyzed their round-trip time (RTT) and failure-rate (FT) under real Internet environment [18,19]. Cavallo et al. collected RTT of services at different time points to constitute the time sequence of QoS, and then applied the autoregressive moving average model (ARMA) to predict such time sequence [20]. Godse et al. further gave the predication of four QoS indexes (RTT, throughput, accessibility and availability), and obtained QoS evaluation by weighting the predicated value [21]. Zhang et al. took into account the social and economic properties of service ecosystem and proposed the people-service-workflow network (PSWN) model [22]. Wu Wenjun and Li Wei et al. investigated the evaluation methods of group software, and analyzed TopCoder and AppStore [23].

**(2) Evolution and Analysis**

In order to improve the undersatanding of service ecosystem, part of scholars analyze the service ecosystem from the prospective of system evolution. Alistair Barros et al. defined five main roles in Web service ecosystem, thus to discuss the provision, discovery and choreography, interaction, quality management, coordination and other key problems of services [24]. Moore pointed out that enterprises play different roles and occupy different ecological niches in service ecosystem depending on their own resources and abilities [15]. Sawatani et al. believed that the service ecosystem combined the self-organizing characteristics of complex system and the coevolution characteristics of ecological system, and owned strong adaptive capacity to the change of the surrounding environment and internal structure [25]. Villalba et al. designed the multi-agent-based simulation model to analyze the features of service ecosystem, including self-organization, self-adaptability and continuous evolvability, etc[26, 27]. Mostafa et al. modeled each service into the independent Service Agent and defined the service composition process as the self-organization collaboration among service agents [28].

**(3) Intervention and Optimization**

In fact, the status of service ecosystem will directly decide the quality of service provision. Hence, it is very important to guide and optimize the evolutionary process of service ecosystem. Some researches used the reinforcement learning method to deal with the dynamics and uncertainty of the internet environment and obtain the optimized service composition [29-32]. Part of study changed the optimization problem of service network into the graph search problem, and the shortest path method is utilized to obtain the optimal solution in the service network [33, 34]. Some study started to introduce the system control concept to the study of service ecosystem. Robin Fischer, Ulrich Scholten, et al. provided a kind of feedback control-based service ecosystem frame to support the control of service provider and the management of service operator [35]. Diao proposed applying the control theory to the management of service system, and achieving the dispatching and management of service by monitoring the service quality [36].

B. *Domain integration of service ecosystem*

In recent years, the service ecosystem theory is also concerned by industrial circle. Both traditional industries and emerging industries are devoted to constructing the service ecosystem to remain their competitive advantages. The typical cases are shown below.

**(1) Software service ecosystem**

The software service ecosystem refers to the complex system formed by the software enterprises through the vertical labor division and horizontal integration of software industry. It opens its software product line and allows the upstream and downstream enterprises, external developers, open source community, even the users to participate in the development and maintenance of software and accelerate [37]. Mohamed et al. proposed a kind of Microservice reference frame based on the autonomic computing, which can reduce the management and evolution cost of large-scale Microservice system by the self-adaptability method [38]. Raji et al. proposed a kind of service evolution modeling method to reduce high cost and possibe errors in the evolution process of microservice system, thus helping the developers effectively manage the Microservice update, framework evolution, deployment environment change, etc. [39].

**(2) Cloud manufacturing service ecosystem**

In order to adapt the network manufacturing trends in the future, more and more enterprises encapsulate their respective distributed resources into the Web service to form the manufacturing service ecosystem in cloud computation platform [40]. Cloud manufacturing service platform integrates and shares diverse and distributed manufacturing resources to cover the whole product development life cycle. The cloud users are allowed to retrieve, purchase and use the manufacturing service as per demand on the platform . It can bring many strengths for manufacturing enterprises to coordinate the cross-border and distributed task [41].

**(4) O2O life service ecosystem**

Online to Offline (O2O) life service ecosystem is a kind of commercial element integration mode, which relies on online ecological engine to drive offline life services by utilizing the mobile internet technology [42]. The O2O life service ecosystem hopes to cover all aspects of people's daily lives, including food, clothing, housing, entertainment, entertainment, etc. After these daily life service resources are redesigned and reorganized to form a closed loop of user consumption, the O2O life service ecosystem can maintain a long-term competitive advantage. The service providers can publish various service information on the Internet at any time, while the service consumers can locate and enjoy the service through direct and real-time inquiry. The personalized service recommendations can make full use of the originally idle service resources, effectively improving the operating efficiency of traditional service industry [43].

In a word, service ecosystem is a continuously-evolved society-technology-economy system. However, current related research still lacks a holistic, systematic performance evaluation model of service ecosystem, and it is difficult for existing methods to reveal the the laws behind the evolution of service ecosystem. In order to face this challenge, this paper proposes a value entropy model of service ecosystem from the perspective of value network, including entropy measurement, value analysis and operation strategy, so as to provide a new technical means for the analysis and intervention of service ecosystem.

III. THE VALUE ENTROPY MODEL OF SERVICE ECOSYSTEM

The service ecosystem is a highly dynamic value generation network, in which the niche of each service node is formed in the process of long-term competition and cooperation. This section proposes the value entropy model of service ecosystem from the prospective that value generation depends on supply and demand matchingof .

A. *Entropy measurement of service ecosystem*

Due to the change of external environment, knowledge acquisition method and other factors, the customers' value demand constantly evolves, which drives the constant evolution of service ecosystem. The value creation process in service ecosystem is shown in Fig.2, in which different colored circles in the value network represent the service nodes at different ecological niches. Service nodes at different ecological niches have different service capabilities and service attributes. When the service demands from external environment are obtained, the service nodes at different ecological niches can work together to jointly create value for customers as per certain value transfer sequence and value distribution rules.

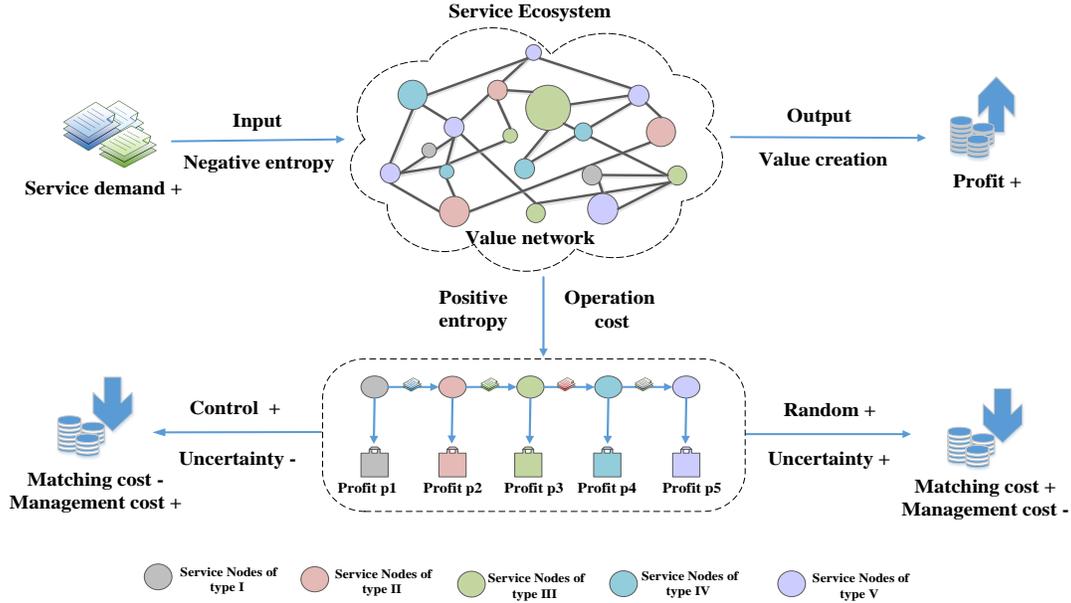

Fig.2 The relationship between service ecosystem and entropy

The measurement of service ecosystem is used to analyze the value creation ability of service ecosystem. However, the concept of value is rather vague and has different definitions in different fields, which makes it difficult to perform quantitative performance evaluation of service ecosystem. Inspired by the application of Entropy concept in information theory, ecology, etc., we proposes the value entropy model to measure the collaborative orderliness of value network, and then to evaluate the value generation efficiency of service ecosystem. As shown in Fig.2, there are management costs (maintaining a certain degree of orderliness of service nodes) and matching costs (finding suitable service nodes) for the operation of service ecosystem, which will continuously generate positive entropy and reduce the orderliness of value network. In order to maintain the continuous operation of ecosystem, it is necessary to continuously input the demands with negative entropy, and generate value through the value network, thereby maintaining the orderliness of value network.

Based on the traditional definition of entropy, the entropy of service ecosystem can be defined as:

$$H_i = -\sum_{j=1}^{n} p_j \log_2 p_j$$

Constraints: $p_j = \dfrac{N_j}{N_{total}}$, and $\sum_{j=1}^{n} p_j = 1$ (1)

In which, $p_j$ represents the distribution probability of service nodes of the $i$-th type, $N_j$ is the number of service nodes of the $j$-th category, and $N_{total}$ is the total number of service nodes in system. In the actual application of Entropy Model, the niche division criteria of service nodes are mainly based on their value creation efficiency. The value creation efficiency of a node can be defined as:

$$E_r = \dfrac{g_r}{c_r} \quad (2)$$

in which $g_r$ is the amount of value created by node $r$ within a certain period, and $c_r$ is amount of value consumed by node $r$ within a certain period of time.

Based on formula (1) and its constraint condition, Lagrange Multiplier Method is adopted to construct formula (3):

$$L(p,\lambda) = -\sum_{j=1}^{n} p_j \log_2 p_j - \lambda(\sum_{j=1}^{n} p_j - 1) \quad (3)$$

The derivative of formula (3) can be obtained as follows:

$$\dfrac{\partial L(p,\lambda)}{\partial p} = -\log_2 p_j - \dfrac{1}{\ln 2} - \lambda = 0$$

$$\Rightarrow p_j = 2^{-\lambda - \frac{1}{\ln 2}}, \; j=1,2,\cdots,n$$

The constraint condition of formula (1) is used to calculate the probability when obtaining the maximum entropy:

$$\sum_{j=1}^{n} p_j = n 2^{-\lambda - \frac{1}{\ln 2}} = 1$$

$$\Rightarrow 2^{-\lambda - \frac{1}{\ln 2}} = \dfrac{1}{n}$$

$$\Rightarrow p_1 = p_2 = p_3 = \cdots p_n = \dfrac{1}{n}$$

Then the maximum entropy value is calculated as:

$$H_{max} = -p_1 \log_2 p_1 - p_2 \log_2 p_2 - \cdots - p_n \log_2 p_n = \log_2 n \quad (4)$$

Because its derivation $H_{max}' = \dfrac{1}{n \ln 2} > 0$, $H_{max}$ is the monotone increasing function, in which $n$ is the number of categories of service node. It indicates the stronger the ecological diversity is, the bigger the entropy value is and the more disordered the ecosystem is. The scale of service ecosystem can decide the upper limit of ecological diversity, thus influencing the changes in entropy.

As a result, if the negative entropy input is less than the positive entropy generated, the entropy value of the system will increase, leading to an increase in the uncertainty of supply and demand matching. Otherwise, the entropy value of the system decreases, resulting in a reduction in the uncertainty of supply and demand matching.

## B. *Value Analysis of service ecosystem*

This sub-section discusses the relationship between entropy and value benefit of service ecosystem. As shown in Fig.3, there are three service ecosystems adopting different operating modes: Fully Controlled Mode, Fully Random Mode and Partially Controlled Mode. The circles represent service nodes in the ecosystem, and the numbers represent the value creation efficiency of the node. The nodes in each dashed box are in the same niche. The internal operation of service ecosystem is based on the niche unit, which consists of two steps: node management and node matching. The node management first divides the nodes into different niches, then sorts the nodes within the niche. As a result, the nodes within the niche are ordered and the niches are out of order. Based on node management, the node matching starts to work. Taking the partially controlled mode as an example, the supply and demand matching needs to first find a suitable niche, and then find the suitable node from the ordered node sequence in the niche.

Here, management cost is expressed as the product of node management time complexity and system unit time cost, and matching cost is expressed as the product of node matching time complexity and system unit time cost. According to the commonly used sorting and seearching algorithms, node management time complexity and node matching time complexity can be set as O ($n\log_2 n$) and O (n), respectively.

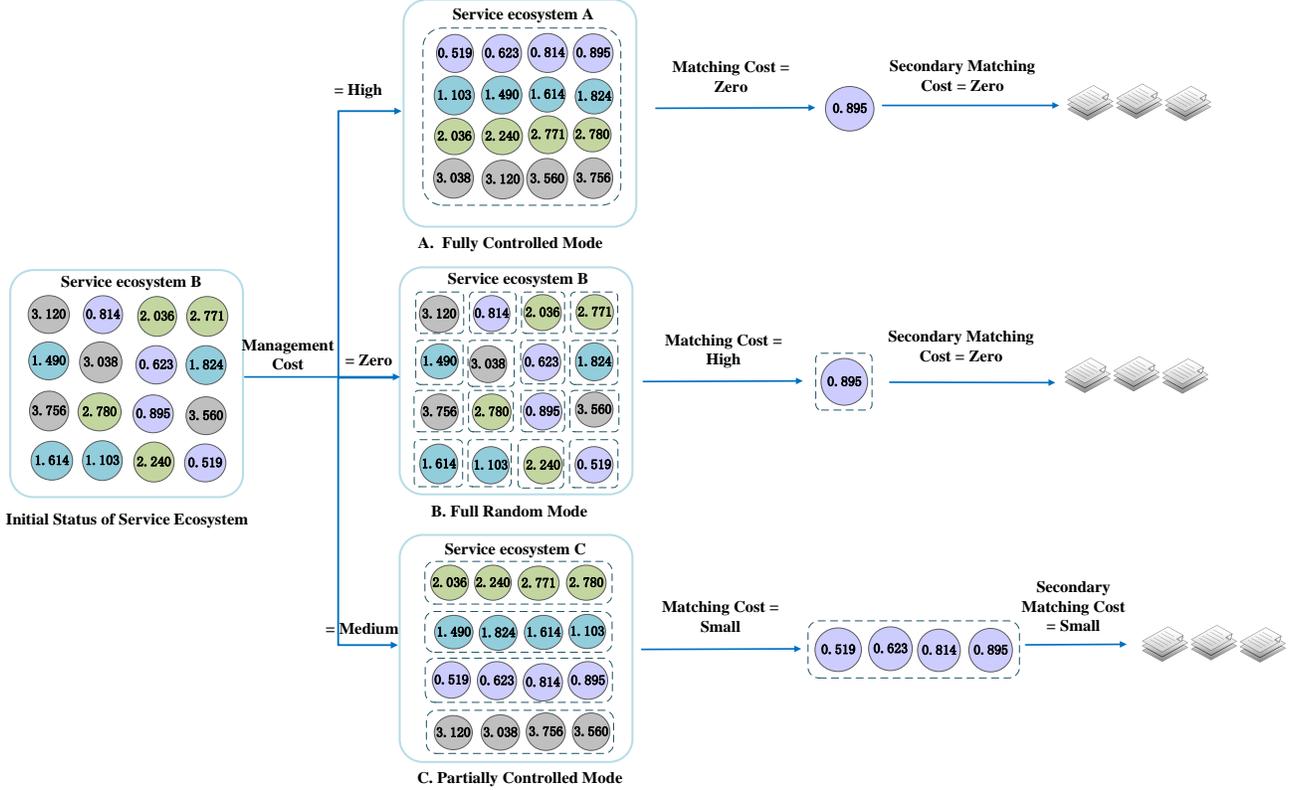

Fig.3 Cost Comparison of three operation strategies of service ecosystem

Considering the scale of service ecosystem, management cost and matching cost can be expressed as follows:

$$c_1 = k \cdot \frac{N}{m} \log_2 \frac{N}{m} \quad (5)$$

$$c_2 = k \cdot m \log_2 m \quad (6)$$

Where $k$ ($k > 0$) is the cost coefficient per unit time of the system, $N$ is the system size, and $m$ is the number of niches, that is, the number of node classifications. Therefore, the operating cost ($c$) and actual value benefit ($v$) of service ecosystem can be expressed as:

$$c = c_1 + c_2 = k \cdot \left( \frac{N}{m} \log_2 \frac{N}{m} + m \log_2 m \right) \quad (7)$$

$$v = \sum_{j=1}^{n} g_j - c \quad (8)$$

According to the derivation of formula(7), we can get the extreme points of costand and then derive the extreme point of value benefit.:

$$c' = k \left( \log_2 m - \frac{N}{m^2} \log_2 \frac{N}{m} + \frac{1}{\ln 2} - \frac{N}{m^2} \frac{1}{\ln 2} \right) = 0$$

$$\Rightarrow m_l = \sqrt{N} \quad //\text{the extreme point}$$

$$\Rightarrow c_{1-\min} = c_{2-\min} = k \cdot \sqrt{N} \log_2 \sqrt{N}$$

$$\Rightarrow c_{\min} = 2k \cdot \sqrt{N} \log_2 \sqrt{N}$$

$$\Rightarrow v_{\max} = \sum_{j=1}^{n} g_j - c_{\min} = \sum_{j=1}^{n} g_j - 2k \cdot \sqrt{N} \log_2 \sqrt{N}$$

The running cost takes the minimum value at the extreme point ($m_l$). On the left side of the extreme point, the cost ($c$) monotonically decreases; on the right side of the extreme point, the cost ($c$) monotonically increases. In contrast, the value benefit ($v$) takes the maximum value at the extreme point.

The relationship between ecological diversity and actual value benefit is shown in Fig.4. It can be known that too high or too low entropy value is not conducive to the value creation of service ecosystem. When the ecosystem reaches the optimal entropy value, the management cost and matching cost reach the equilibrium point, the operation cost is the lowest, and the actual value benefit is the largest. The optimal value entropy can be expressed as follows:

$$H_b = \log_2 m_l = \log_2 \sqrt{N} \qquad (9)$$

Here, we take the three service ecosystems in Fig.4 as examples to illustrate the nonlinear relationship between entropy and value benefit.

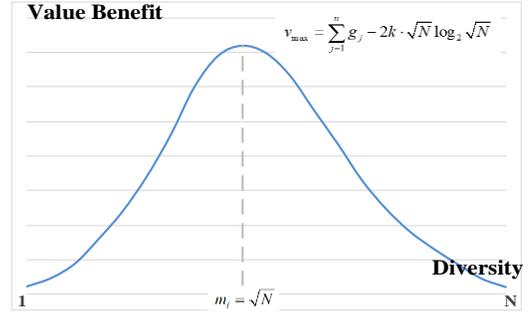

Fig.4 The relationship between ecological diversity and actual value gain

TABLE.1
THE COMPARISON OF SYSTEM INDICATIONS OF THREE CASES IN FIG.4

| Market Env / Opeartion Mode | Demand sequence | Single niche | Entropy | Cost | Value benefit |
|---|---|---|---|---|---|
| Service ecosystem A (fully controlled mode) | the value is $V$ | A single niche (only one node type) | $H_A = \log_2 1 = 0$ | management cost >> matching cost $c_A = k(16\log_2 16 + 1\log_2 1) = 64k$ | $v_A = V - c_A = V - 64k$ |
| Service ecosystem B (fully random mode) | the value is $V$ | Each node is in an independent niche ($N$ node types) | $H_B = \log_2 16 = 4$ | management cost << matching cost $c_B = k(1\log_2 1 + 16\log_2 16) = 64k$ | $v_B = V - c_B = V - 64k$ |
| Service ecosystem C (partially controlled mode) | the value is $V$ | All nodes are divided into $m$ niches, and keeps the order of nodes in each niche. ($m$ node types) | $H_C = \log_2 4 = 2$ | management costs and matching costs are more balanced. $c_C = k(4\log_2 4 + 4\log_2 4) = 16k$ | $v_C = V - c_C = V - 16k$ |

Since $k > 0$, the entropy values and value benefit in the three modes are compared as follows:

$$H_A < H_C < H_B, \text{ and } v_C > v_B = v_A$$

Among them, the entropy value of service ecosystem A is the smallest, and its unnecessary management costs are paid for the strict internal control; the entropy value of service ecosystem B is the largest, its internal collaboration is too disordered and the matching cost is high; the entropy value of service ecosystem C is closest to the optimal value entropy, its management cost and matching cost are the most balanced and it has the largest actual value benefit.

C. *Operation strategy of service ecosystem*

As a supply-demand matching system, the evolution process of service ecosystem is not only affected by the supply-side management mode, but also by the demand-side environment. In practice, there are two kinds of typical demand secnarios: a mature market environment ($D_M$) and a emerging market environment ($D_E$). In a mature market environment, the market potential has been fully developed and the number of demands has remained stable for a long time. In the emerging market environment, the market potential has not been fully developed, and the number of demands may show explosive growth.

The operating cost of service ecosystem consists of management cost and matching cost. For different operating modes, the decisive elements of their cost composition are different. For the control-dominated mode, management cost plays a decisive role, which is proportional to the amount of service nodes; while for the random-dominated mode, matching cost plays a decisive role, which is proportional to the amount of demands. Therefore, the cost of control-dominated mode ($C_a$) and random-dominated mode ($C_b$) can be expressed as formula 10 and formula 11:

$$c_A = k \frac{N}{m_A} \log_2 \frac{N}{m_A} \qquad (10)$$

$$c_B = k m_B \log_2 m_B * d \qquad (11)$$

Where $k$ ($k > 0$) is the cost coefficient per unit time of the system, $N$ is the number of nodes (i.e. system size), $m_a$ is the number of niches in the control-dominated mode, and $m_b$ is the number of niches in the random-dominated mode, $d$ is the amount of demands in the market environment.

TABLE.2
THE COST OF TWO OPERATION MODES IN DIFFERENT MARKET ENVIRONMENTS

| Market Env / Opeartion Mode | Stable Demand Sequence | Explosive Demand Sequence |
|---|---|---|
| Control dominated Mode | $C_a$ (large) | $C_a$ (small) |
| Random dominated Mode | $C_b$ (small) | $C_b$ (large) |

Based on the above conclusions, Table 2 shows the cost representation of different operating modes in different market environments.

(1) In a mature market environment, the system still maintains a considerable scale, but the number of needs to be processed is not large. Based on formula (10) and (11), the cost of the control-dominated mode ($C_a$) is fixed and proportional to system size; the cost of the random-dominated mode ($C_b$) remains low. As a result, the cost of the control-dominated mode is larger than that of the random-dominated mode ($C_a > C_b$). So, in a mature market environment, a random-dominated mode with strong ecological diversity has an advantag, that is, the larger the entropy value, the higher the value benefit.

(2) In a emerging market environment, the system still maintains a considerable scale, but the number of demands to be processed has grown dramatically. Based on formula (10) and (11), the cost of the control-dominated mode ($C_a$) is fixed; the cost of the random-dominated mode ($C_b$) continues to increase sharply as the matching frequency increases. As a result, the cost of the control-dominated mode is smaller than that of the random-dominated mode ($C_a < C_b$). So, in a emerging market environment, a control-dominated mode with weak ecological diversity has an advantage, that is, the lower the entropy value, the higher the value benefit.

Further, we can find the dividing point of the quantity of demand ($d$), so as to make a quantitative distinction between the two market environments. The derivation process is as follows:

$$k \frac{N}{m_A} \log_2 \frac{N}{m_A} = k m_B \log_2 m_B * d$$
$$\Rightarrow d' = \frac{N*(\ln N - \ln m_A)}{m_A * m_B * \ln m_B} \quad (12)$$

As shown in Fig.5, when $d < d'$, the cost of control-dominated mode is higher ($C_a > C_b$) and its value benefit is lower; when $d > d'$, the cost of random-dominated mode is higher ($C_b > C_a$) and its value benefit is lower.

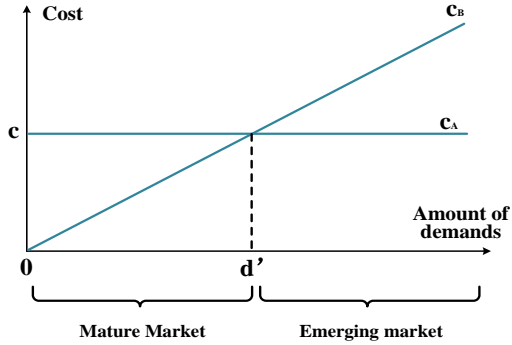

Fig.5 Relationship between model mode and market environment

According to formula (12), when the scale of service ecosystem is expanded, $d'$ will move to the right, and the range in which the random mode is dominant will become larger; when the scale of the service ecosystem is reduced, $d'$ will move to the left, and the scope of the control mode is dominant will become larger. The above analysis results provide basis for studying the optimized operation strategy of service ecosystem. We can draw the following related conclusions based on the value entropy model:

(1) For a given market environment, there is an optimal entropy value that maximizes the value benefit of the service ecosystem. Too high or low entropy is not conducive to the creation of actual system value benefit.
(2) Under the condition of mature market environment, the random-dominated mode with higher entropy value has more advantages and higher returns.
(3) Under the condition of emerginge market environment, the control-dominated mode with lower entropy value has more advantages and higher returns.

## IV. DESIGN OF COMPUTATIONAL EXPERIMENT SYSTEM

In order to verify the applicability of the value entropy model, the corresponding computational experiment system is constructed as the artificial society laboratory. Borrowing the idea of service bridge [44], the evolution of service ecosystem can be abstracted as a continuous matching process between the supply side and demand side. Based on the concept, related design details are divided into three parts: design of supply side, design of demand side, and design of system operation.

### A. Design of Supply-side

The agent is an entity with characteristics of autonomy, society, reaction, and pre-action. Service nodes in the service ecosystem have similar characteristics, such as interconnection rather than isolation, autonomy rather than obedience, etc. Therefore, the agent becomes a natural metaphor for the active entities of the ecosystem. In the computational experiment system, the supply-side agent stands for the service nodes offering goods or services. They are active and dynamic, serving as the active behavior entity in system environment.

All supply-side agents search their own specific orders (e.g. primary node ->primary order, secondary node->secondary order, and third-level node ->third-level order) in the environment and consume certain capital in the searching process. After acquiring orders, their own capital will increase accordingly and produce secondary orders for downstream nodes. When their capital reaches the reproduction threshold, genetic evolution is conducted to produce new child agents of the same kind. When their capital is smaller than their death threshold, they die and disappear.

The survival of the fittest among service nodes are key factors driving the evolution of service ecosystem. In the intense competition among service nodes, those nodes that are not competitive are likely to be eliminated. In order to survive in the ecosystem, service nodes must improve their decision-making and behavioral skills through a variety of learning methods. The evolution process of individual node is the result of the combined effects of individual learning, organizational learning and social learning.

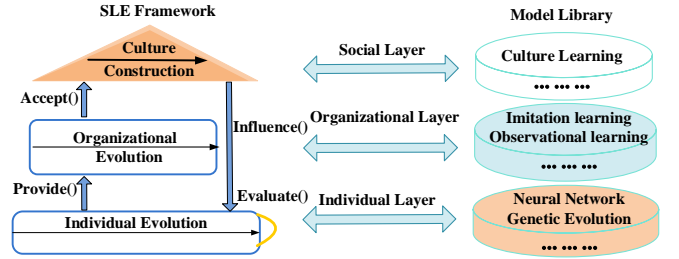

Fig.6 SLE Modeling Framework

Based on the work in [45], the SLE framework is used to describe the characteristics of the supply side agents. As shown in Fig.6, the SLE framework is composed of two parts: the left column indicates three modeling layers and the right column indicates the implement models adopted by each layer. There is a feedback loop among the three modeling layers: the modeling of individual evolution is at the bottom layer, which simulates the genetic evolution phenomenon of individual node in service ecosystem; the modeling of organizational evolution is at the intermediary layer, which simulates the imitation and observational learning among service communities; the modeling of social evolution is at the top layer where the knowledge of some elites can be extracted as culture, which simulates the accelerated evolution of the whole ecosystem promoted by culture.

The SLE is a customizable modeling framework. Depending on the specific needs, the models and techniques required for each layer can be selected and adopted from the corresponding model library. Model elements in the model library can be added, deleted, and modified as needed. The implementation details of each layer are shown as follows:

- **Individual evolution layer**: The bottom layer is the micro level, which is used to simulate the independent evolution of individual service nodes in the real world. According to the rule of survival of the fittest, each individual node needs to continuously improve its own ability in order to survive in the fierce market competition. The evolutionary models commonly used here include genetic algorithms, reinforcement learning, neural networks, and so on.
- **Organizational evolution layer**: The middle layer is the organizational level, which is mainly used to simulate the cooperation between service nodes to enhance the competitiveness. In the real world, market competition has evolved from the competition between single nodes to the competition between groups. The evolutionary models commonly used here include observational learning, imitation learning, and so on. Different evolution mechanisms can lead to different outcomes.
- **Social evolution layer**: The top layer is the cultural level, which is mainly used to simulate the impact of elite culture on individual evolution in society. In the real world, some elites with excellent knowledge will gradually emerge from the group because of their excellent performance. Then, their knowledge can be extracted into culture, and it can affect the individual evolution at the micro level. For example, the operation mode of service ecosystem (random mode or control mode) can accelerate or hinder the development of many single nodes in different scenarios.

*B. Design of Demand Side*

In experiment system, the demand-side elements (i.e. order) form the system environment module together, which is regarded as a container for all supply-side agents. The position of each order is fixed in its entire lifecycle. If one order is processed by some agents or its lifecycle is over, it will disappear. After some fixed time cycles, new orders will emerge according to the designed order generation model, including order amount, order category (e.g. primary-level order, second-level order, third-level order), order distribution, profit value of order unit, etc. Thus, all kinds of market fluctuation trend can be simulated, such as mature market environment and emerging market environment.

In experiment system, each order has a certain complexity, that is, the order needs to go through several links to be processed. Fig.7 shows the virtual "food chain" including three types of service nodes. Primary orders are the source of value benefit for all nodes, which requires three steps to complete. At first, the primary order is disposed by first-level service node and generates the secondary order, and then the secondary order will be handled by secondary node to produce the third-level order. Until the third-level order is processed by third-level nodes, all the service nodes in the relevant links can obtain the corresponding share of profits. If the third-level order is not disposed by specific service nodes within the period, early profits generated by the primary order will also be invalid.

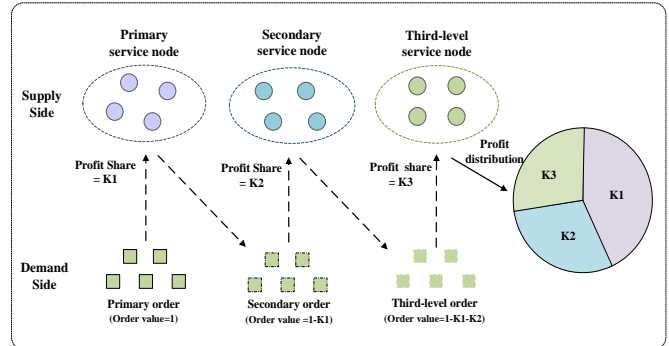

Fig.7 The "Food-chain" Relation in Service Ecosystem

Here, the demand-side characteristics of a particular domain can be descripted by Formula (13):

$Demand\_Char =< Trends, Volume, Location, Category, QoS\ Preference >$ (13)

Based on our existing domain knowledge, the content of each element can be described as follows:

- ***Trends.*** It represents the macro market characteristics of the analyzed domain. In real world, the fluctuation of market trends conforms to a certain rule. According to the magnitude of fluctuation, the trends can be categorized as stable market trends and fluctuated market trends.
- ***Volume.*** It represents the potential market size of the domain analyzed. In actual environment, the market size varies a lot among domains, which is determined by the user's purchase frequency and product unit price. For example, E-commerce service belongs to high-frequency & high price domain, which has a large potential market size.
- ***Location.*** It represents the geographical location of the domain analyzed. Because of differences in economic development levels, customer consumption habits, etc., the market characteristics vary greatly among regions, which may be reflected in the number of orders, type of orders, unit price of orders, etc.
- ***Category.*** It represents the diversity of the service demands. Different demands need to be met by different service providers. Even for the same product or service, service providers may vary a lot in price and quality.
- ***Complexity***: It represents the number of links an order needs to be processed. Generally speaking, the complex orders need to be processed by different service chain links in turn. If the initial complexity of an order is 3, the value network needs to complete such order through the cooperative process between three service links.
- ***QoS (Quality of Service) Preference***. It represents the dynamics of user preferences. In the real world, not only the total number of customer demands may change, but the preferences of individual demand may also change. For example, with the development of social economy, customers' consumption will be upgraded, from price preference to quality preference. This change will lead to a reduction in the size of the original market and an increase in the share of emerging markets

## C. Design of System operation

The evolution direction of service ecosystem is determined by the matching process between the supply-side and the demand-side. As shown in Fig.8, a variety of experiment scenarios can be customized by setting and combing the supply-side and demand-side parameters. Without external intervention, experimental systems can be used to simulate the natural evolution of service ecosystem. The evolution result depends mainly on initial conditions and internal mechanisms. If there is external intervention, the experimental system is mainly used to simulate controlled evolution to assess the effectiveness of the intervention. By observing the evolution phenomena of ecosystem in experiment system, it is possible to intuitively find the appropriate intervention strategies.

In the operation scene of our experiment system, two service ecosystems adopting different strategies are constantly playing against each other. The service ecosystem α adopts a control-dominated strategy, and the service ecosystem β adopts a random-dominated strategy. The red symbols represent the service node in service ecosystem α and the blue symbols represent the service node in service ecosystem β. The entire scene is divided into 5 regions: initial area (1 and 4), adjacent area (2 and 5), and emerging area (3). The complexity of orders in the initial region, adjacent region and emerging region is 1, 2, and 3 respectively. They represent the core business, related business, and emerging business of a service ecosystem respectively. Green area represents the order-rich regions, and the depth of the green reflects the intensity of orders.

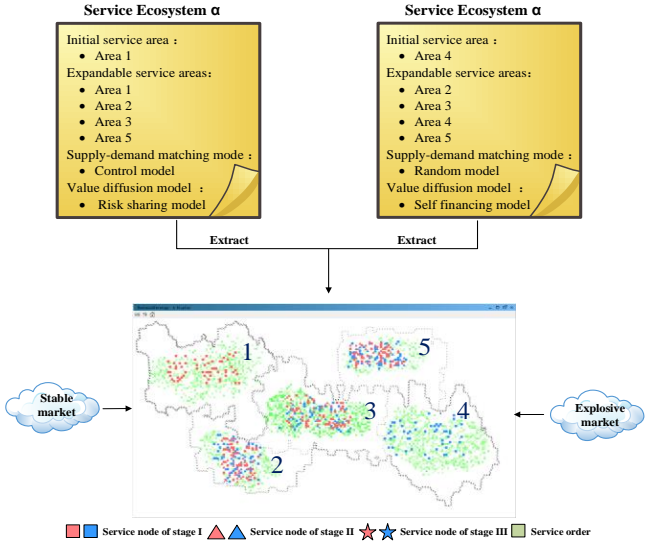

Fig.8 The design of computational experiment system

In the initial state of the experiment, the agents of two ecosystems are respectively distributed in their core business areas: service ecosystem α is in area 1, and service ecosystem β is in area 4. As the scale of the service ecosystem expands, their agents will gradually enter neighboring areas and emerging areas. During this period, three types of agents (primary node, secondary node, and third-level node) will be produced, which are represented by different symbols with different shapes (square, triangle, star). The farther the agent is from its core area, the higher the cost it consumes per unit time. When agents belonging to different ecosystems meet, the one with large capital value can kill the other and possess its value. As more and more agents from different ecosystems enter the same area, the competition between them will become more intense.

In the computational experiment environment, various operation mechanisms of service ecosystem can be evaluated, including some pressure test and boundary test. The purpose of the service operation strategy is to adjust the relationship between different nodes. The performance of different service strategies varies widely. Here, we take two service operation strategies as the experiment objects. The related details are given as follows.

**Option 1:** Control-dominated strategy

The control-dominated strategy will use the virtual hub to coordinate the management of all nodes, corresponding to the strong relationship of the value network. The type of service ecosystem has a strong ability to share risks. After a fixed period, the virtual hub will collect the profits of all nodes and then distribute them to all nodes according to certain rules. In this way, a single node has stronger survivability, and it is easier to go farther from the core area.

**Option 2:** Random-dominated strategy

The random-dominated strategy emphasizes the autonomy of service nodes and the equal cooperation between nodes, which corresponds to the weak relationship of the value network. In the process of value creation, each service node is responsible for its own profits and losses, and there is no risk sharing mechanism. The survival of the fittest among the nodes leads to a stronger adaptability of the entire system. In this way, the viability of a single node is not strong, and it will not easily deviate from the core area.

## V. EXPERIMENT EVALUATION OF SERVICE ECOSYSTEM

In this section, various experiment scenarios are designed to compare the performance of two service ecosystems that adopt different operating strategies. The experimental results will be used to verify the validity of the entropy model.

### A. Initialization of Computational Experiment

**(1) Experimental scene**

Case 1, the overall market demand remains relatively stable only with periodic and small range fluctuation; case 2, the market demand in the emerging area explodes in the 280th cycle.

**(2) Experimental subject**

The experiment set up a competitive game between service ecosystem α and β within the same environment. Service ecosystem α and β adopt control-dominated strategy and the random-dominated strategy respectively.

**(3) Parameter setting**

In the construction of the "New Retail" business ecosystem, Alibaba and Tencent have adopted control-dominated and random-dominated strategies, respectively. So, they are used as the prototypes of service ecosystem in the experiment. The experimental parameters refer to the public operating data of the two companies from 2015 to 2018, which are set as follows:

TABLE 2
PARAMETERS SETTING OF COMPUTATIONAL EXPERIMENT

| System Variable | Experiment Setting |
|---|---|
| **Environment Setting** | |
| Environment size | 250*120 |
| **Agent Setting** | |
| Initial number of primary service nodes | α=12, β=14 |
| Initial capital value | Bounded random within the range of [180,220]. |
| Agent Type | Bounded random within [1,3]. |
| Distance cost | Y=k*x(x>0, x indicates the distance moved.) In area1 and area4, k=1. In area2 and area5, k=1.3. In area3, k=1.7. |
| Operation cost | Bounded random within the range of [3,5] in area1 and area4. Bounded random within the range of [3,7] in area2 and area5. Bounded random within the range of [3,9] in area3. |
| Speed | Bounded random within the range of [1, 5]. |
| Vision range | Bounded random within the range of [1 ,5]. |
| Reproductive threshold | 300 |
| Reproductive Punishment | Y=k*d(x indicates the distance between parent agent and child agent, k=3) |
| Expansion threshold | N=25,V=4000 for area 2 and area 5. N=125,V=15000 for area 3. |
| **Order Setting** | |
| Complexity | Bounded random within the range of [1,3]. |
| Order Type | Bounded random within the range of [1,3]. |
| Order Value | Bounded random within the range of [10,30] when its initial complexity = 1. Bounded random within the range of [50,80] when its initial complexity = 2. Bounded random within the range of [70,100] when its initial complexity = 3. |
| Distribution of order | Orders are distributed randomly in five areas with centers of (59,79)(area1), (85,26)(area2),(125,54)(area3),(157,90) (area4), (180,36) (area5) respectively. |
| The generation rule of order | The market trends are represented by the function Y=N+M*sin(t). In area1 and area4, the reference value of order amount N is set as 200 and the range of fluctuation M is set as25. In case 1, the reference value of order amount N is set as 225 and the range of fluctuation M is set as 30 in area2, area3 and area5, In case 2, the reference value of order amount N is set as 350 in area3 when tick =280, and others are the same as case 1. |
| The profit sharing ratio | The ratio is 6:4 when the initial complexity of orders is 2. The ratio is 4:3:3 when the initial complexity of orders is 3. |

### (4) Evaluation indicators

The performance indicators introduced by the experimental system include entropy value $H$, which is used to measure the degree of disorder of service ecosystem; value consumption $C$, which is used to measure the operating costs of service ecosystem; and system value benefit $V$, which is used to measure the sustainability of service ecosystem.

### B. Case 1: Ecosystem evolution in mature market

The evolution process of service ecosystem in mature marketing environment is shown in Fig.9.

(1) As shown in Fig.10-A, during the initial stage (from Tick=0 to Tick=80), there are only the primary service nodes in both service ecosystems and the nodes are only distributed in their core areas.

(2) As shown in Fig.10-B and Fig.10-C, during the early stage (from Tick=80 to Tick=160), both service ecosystems reach the expansion threshold simultaneously and the nodes enter into the adjacent areas.

(3) As shown in Fig.10-D and Fig.10-E, during the middle stage (from Tick=160 to Tick=320), service ecosystem α reaches the expansion threshold firstly and enter into the emerging service area.

(4) As shown in Fig.10-F, during the later stage (from Tick=320 to Tick=400), both service ecosystems is gradually taking shape. During the evolution, the number of nodes in both systems has been increasing. Because the control-dominated strategy has a risk sharing mechanism, the number of agents in the service ecosystem α is higher in high-risk areas.

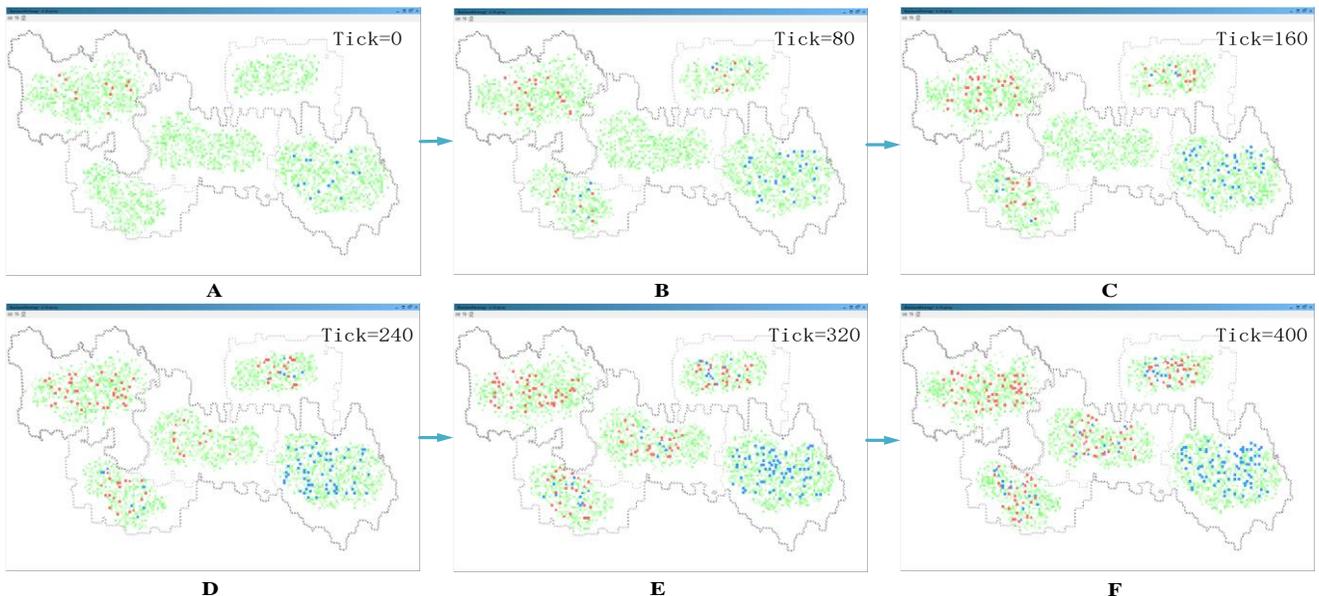

Fig.9 The evolution process of two service ecosystems in Case 1

Fig.10 gives a comparative analysis of the performance indicators of the two service ecosystems in Case 1. Fig.10-A gives the change of order quantity in all areas during the experiment period.

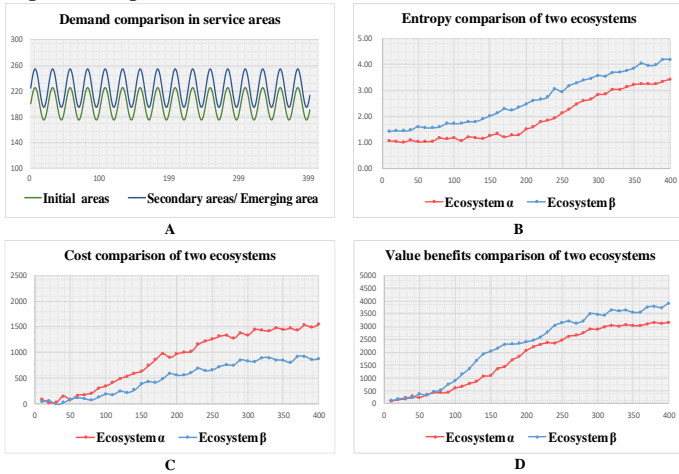

Fig.10 The performance comparison of two ecosystems

(1) Fig.10-B is the comparison of the entropy of two service ecosystems. With the increase in the number of service nodes, the diversity and disorder of the two systems are constantly increasing. Because ecosystem β adopts a random-dominated strategy, its ecological diversity is strong, and its entropy value is continuously greater than that of ecosystem α.

(2) Fig.10-C is the comparison of the cost of two service ecosystems. In a mature market environment, and management costs play a decisive role. Ecosystem α adopts the control-dominated strategy, which is more affected by management costs. During the experiment period, the system cost of ecosystem α and β increase slowly, and the cost value of α is higher than that of β. .

(3) Fig.10-D is the comparison of value benefits of two service ecosystems. In the same competitive environment, the higher the cost, the lower the net profit. So, the service ecosystem β has a higher value benefit in the evolution process.

Based on experiment analysis, it can be seen that in the mature market environment, the service ecosystem α is more orderly, but its value benefits and value growth trend are lower than that of β. The experiment results show that, the random-dominated strategy has better performances when the demand environment is stable. This result is consistent with the second conclusion of entropy model analysis.

## C. Case 2: Ecosystem evolution in emerging market

The evolution process of service ecosystem in emerging market environment is shown in Fig.11.

(1) As shown in Fig.11-A, 11-B and 11-C, the early stage of this group of experiments is roughly the same as Case 1, and the differences are mainly in the middle and late stages of the experiment.

(2) As shown in Fig.11-D and Fig.11-E, during the middle stage (from Tick=160 to Tick=320), service ecosystem α reaches the expansion threshold firstly and enter into the emerging service area. In this area, the market demands show the explosive trend.

(4) As shown in Fig.11-F, during the later stage (from Tick=320 to Tick=400), both service ecosystems is gradually taking shape. The burst of demands causes more nodes in Case 2 than that of Case 1 in the same period. Because service ecosystem α adopts the control-dominated strategy, it has more agents in high-risk areas than ecosystem β.

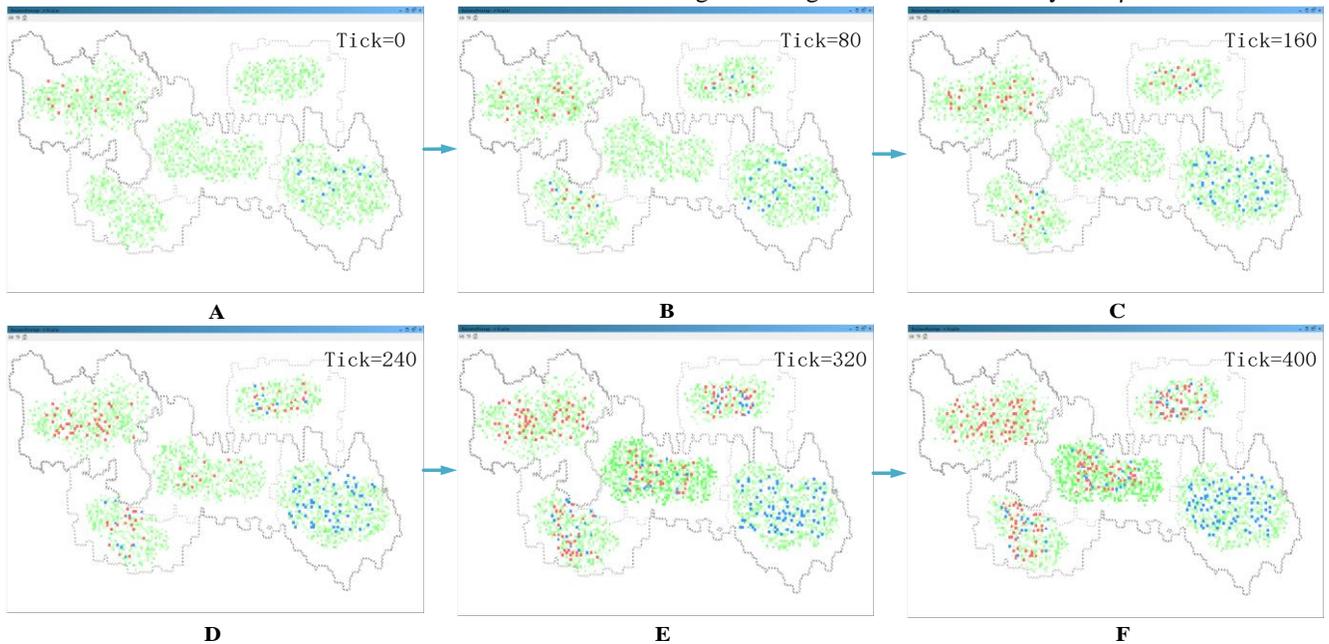

Fig.11 The evolution process of two service ecosystems in Case 2

Fig.12 gives a comparative analysis of the performance indicators of the two service ecosystems in Case 2. Fig.12-A gives the change of the order quantity in all areas during the experiment period, and in the middle and late stage, the demands explode in the emerging areas.

(1) Fig.12-B is the comparison of the entropy of two service ecosystems. The change trend of the entropy curve of the two systems is roughly the same as that of Case 1. But in this experiment, the number of nodes of the two systems is greater, and the ecological diversity is stronger. Therefore, Case 2 has a higher entropy value than Case 1.

(2) Fig.12-C is the comparison of the cost of two service ecosystems. In the early stage of the experiment, demand quantity is stable, and management costs steadily increase with the number of nodes. At this stage, the main share of overall costs is management costs. In the middle and late stages of the experiment, due to the outbreak of demand, matching costs increase significantly. Ecosystemβ adopts random-dominated strategy, which is more affected by matching costs. This leads to the total cost of ecosystem β overtaking ecosystem α in the later stage of the experiment.

(3) Fig.12-D is the comparison of value benefits of two service ecosystems. The cost of service ecosystem is inversely proportional to value benefit. In the latter part of Case 2, the cost of ecosystem β surged and surpassed ecosystem α. So, the value of ecosystem α finally surpassed that of ecosystem β.

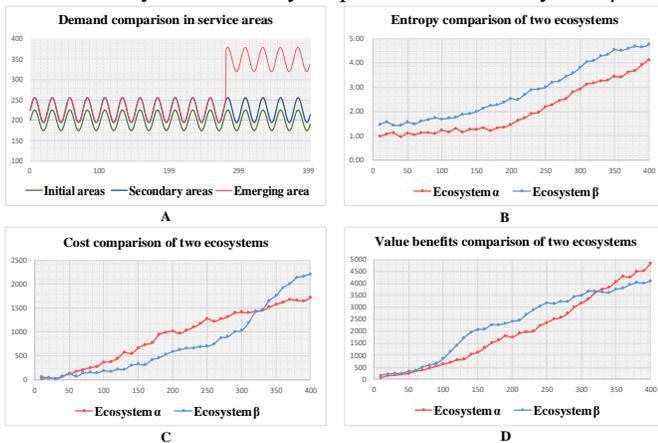

Fig.12 The performance comparison of two ecosystems

Based on experiment analysis, it can be seen that in the emerging market environment, the value benefits and value growth trend of service ecosystem α are also greater than that of β. The experiment results show that, the control-dominated strategy has better performances when the demand environment is explosive. This result is consistent with the third conclusion of entropy model analysis.

In Case 1, the service ecosystem α with lower entropy has lower value benefits. In Case 2, the service ecosystem β with higher entropy has lower value benefits. This shows that too high or too low entropy is not conducive to creating value. The system whose entropy is closer to the optimal entropy value has higher value benefits, which is consistent with the first conclusion of the value entropy model.

## VI. DISCUSSION

This section will compare the differences between Alibaba and Tencent in the construction of the service ecosystem to prove the validity of the experimental results. The relevant data comes from their financial reports, official website and related service data in the APP Store.

As shown in Fig.13 and 14, the construction processes of service ecosystem of the two Internet enterprises can be approximately divided into three stages. In the first two phases, both companies focused on their core areas and related business areas. In the third phase, New retail, Mobile payment and other emerging areas (e.g. Cloud computing and IoT platforms) have brought a new round of development opportunities for the two enterprises.

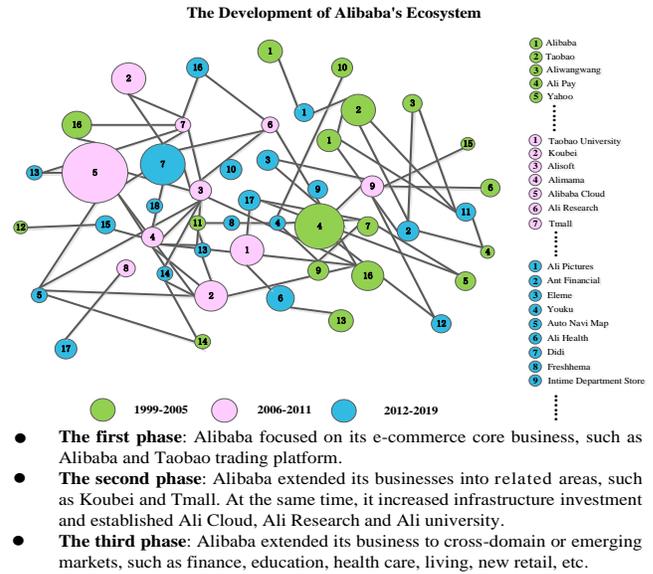

- **The first phase**: Alibaba focused on its e-commerce core business, such as Alibaba and Taobao trading platform.
- **The second phase**: Alibaba extended its businesses into related areas, such as Koubei and Tmall. At the same time, it increased infrastructure investment and established Ali Cloud, Ali Research and Ali university.
- **The third phase**: Alibaba extended its business to cross-domain or emerging markets, such as finance, education, health care, living, new retail, etc.

Fig.13 The construction process of Alibaba's service ecosystem

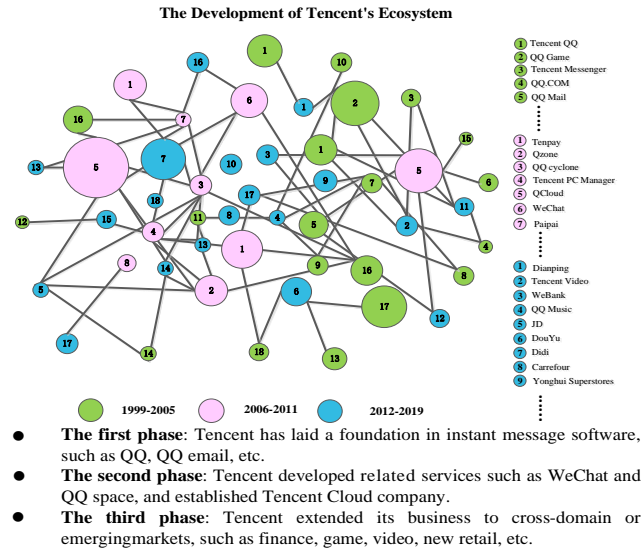

- **The first phase**: Tencent has laid a foundation in instant message software, such as QQ, QQ email, etc.
- **The second phase**: Tencent developed related services such as WeChat and QQ space, and established Tencent Cloud company.
- **The third phase**: Tencent extended its business to cross-domain or emergingmarkets, such as finance, game, video, new retail, etc.

Fig.14 The construction process of Tencent's service ecosystem

The different development strategies mentioned in the experiment are also reflected in the operation of the two Internet companies. Alibaba's strategy is control-dominated, and it emphasizes the full control over nodes in the ecosystem. It takes e-commerce business as the core of the whole ecosystem and all other businesses are built around this core, including finance, logistics, cloud computing and other related fields. In order to ensure the deep convergence of its core area and emerging areas, Alibaba either set up the company itself or bought other companies wholly, such as Cainiao Logistics, Qunar, etc. This strategy has a relatively strong execution power and can continuously invest in emerging areas.

On the other hand, Tencent has adopted a random-dominated strategy, emphasizing itself as the ecosystem's infrastructure. It empowers related enterprises with the resources needed to form a loose community of interests. Tencent's advantage lies in online traffic. It enters areas where it is not good at by investing in shares, such as E-commerce, Sharing economy, O2O life service, and so on. The advantage of this strategy is less investment and relatively low risk. However, it is difficult to make long-term investment in some areas with uncertain prospects, and some valuable opportunities may be missed.

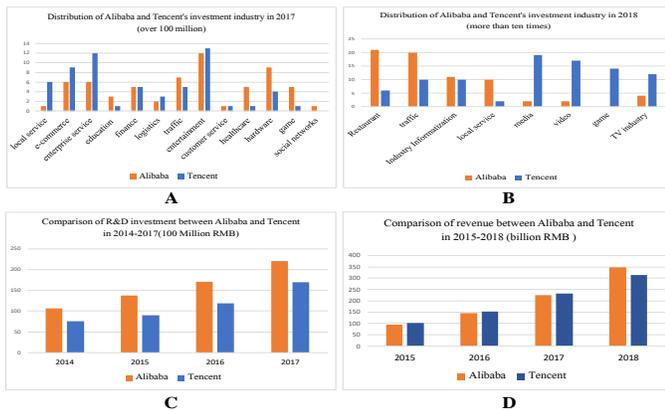

Fig.15 The comparison of financial data between Alibaba and Tencent

As shown in Fig.15-A and 15-B, the ecosystems of Alibaba and Tencent have shown an increasing tendency to overlap with each other in business areas. Alibaba's development strategy requires more investment in the early stage. As shown in Fig.15-C, since 2014, Alibaba's investment has continued to be higher than that of Tencent. Figure 15-D shows the revenue comparison between Alibaba and Tencent from 2015 to 2018. Before 2017, the emerging markets have not yet been broken, and market demand is relatively stable. Therefore, Tencent has an advantage in terms of revenue. Subsequently, cloud computing and mobile payment businesses experienced explosive growth. In 2018, Alibaba's revenue exceeded Tencent's, and its control-dominated development strategy played a huge role in it, which is consistent with the analysis results of our entropy model.

## VII. Conclusion

As a product of service-based economy and software service technologies, service ecosystem is a complex socio-technical system. In order to better study and manage the complex and dynamic relationships between the service nodes in the ecosystem, this paper proposes a value entropy model of service ecosystem from the perspective of value network, including entropy measurement, value analysis and operation strategy, so as to provide a new technical means for the analysis and intervention of service ecosystem. The service ecosystem theory has been widely used in different fields such as manufacturing, e-commerce, and information services. The value entropy model is universal because it does not depend on specific domain attributes. So, it can make unified and reasonable evaluation for different types of service ecosystems. The above work can provide new research ideas and tools for the evolutionary analysis of service ecosystem and the optimal governance of service ecosystems.

The purpose of interpreting phenomena is to predict, while the purpose of prediction is to control. In order to achieve a controlled evolution of service ecosystem, there are many areas that need further research, including how to realize the precise control of evolution process, how to implement minimum cost control, how to choose the best control point, and so on. In the field of mobile Internet service ecology, there are many competition cases where small companies beat big companies, such as the impact of ByteDance on Tencent and the challenge of Meituan on Alibaba. In our entropy model, the measure of collaborative disorder does not depend on the system scale, which provides a basis for studying the hidden drivers of the rise of small and medium-sized ecosystems in fierce competition. In the future, we will use the continuously optimized entropy model to analyze the evolution of service ecosystems in different fields and different sizes. Furthermore, we can reveal the explicit and implicit driving factors among them, so as to provide the optimal evolution path of the service ecosystem in the corresponding context.

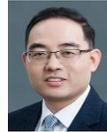

**Xiao Xue,** born in 1979. Professor with the School of Computer Software, College of Intelligence and Computing, Tianjin University. Also adjunct professor in the School of Computer Science and Technology, Henan Polytechnic University. His main research interests include service computing, computational experiment, Internet of things, etc.

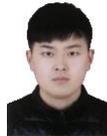

**Zhaojie Chen,** born in 1994. Graduate student in the School of Computer Science and Technology, Henan Polytechnic University. His current research interests include service computing.

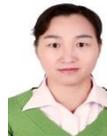

**Shufang Wang,** born in 1979. Associate professor and master supervisor in the School of Geographic and Environmental Sciences, Tianjin Normal University. Her main research interests include human geography and computational experiment, etc.

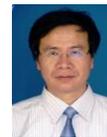

**Zhiyong Feng**, born in 1965. Professor and PHD supervisor in the School of Computer Software, Tianjin University. His main research interests include service computing, software engineering, Internet of things, etc

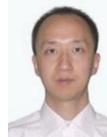

**Yucong Duan,** born in 1977. Professor and director of Computer Science Department at Hainan University. His research interests include Service Computing, Knowledge Graph, Big Data, etc.

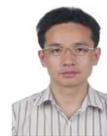

**Zhangbing Zhou,** born in 1975. Professor with the China University of Geosciences, Beijing, China, and as an Adjunct Professor at TELECOM SudParis, Evry, France. He has authored over 100 referred papers. His research interests include process-aware information system and sensor network middleware.